\documentclass[letterpaper]{article} 
\usepackage[submission]{aaai24}  
\usepackage{times}  
\usepackage{helvet}  
\usepackage{courier}  
\usepackage[hyphens]{url}  
\usepackage{graphicx} 
\usepackage{natbib}  
\usepackage{caption} 
\usepackage{algorithm}
\usepackage{algorithmic}

\usepackage{newfloat}
\usepackage{listings}

\usepackage{makecell}

\usepackage{color}

\DeclareCaptionStyle{ruled}{labelfont=normalfont,labelsep=colon,strut=off} 
\floatstyle{ruled}
\newfloat{listing}{tb}{lst}{}
\floatname{listing}{Listing}
\title{Landscape of Generative AI in Global News: Topics, Sentiments, and Spatiotemporal Analysis} 
\author{
    Lu Xian \textsuperscript{\rm 1}\equalcontrib,
    Lingyao Li \textsuperscript{\rm 1}\equalcontrib,
    Yiwei Xu \textsuperscript{\rm 2},
    Ben Zefeng Zhang \textsuperscript{\rm 1},
    Libby Hemphill \textsuperscript{\rm 1}
}
\affiliations{
    \textsuperscript{\rm 1}School of Information, University of Michigan\\
    \textsuperscript{\rm 2}Information School, University of Washington\\

    xianl@umich.edu, lingyaol@umich.edu, yiweixu@uw.edu, bzfzhang@umich.edu, libbyh@umich.edu \\
}

{}

\usepackage{bibentry}

\begin{document}

\maketitle

\begin{abstract}
Generative AI has exhibited considerable potential to transform various industries and public life. The role of news media coverage of generative AI is pivotal in shaping public perceptions and judgments about this significant technological innovation. This paper provides in-depth analysis and rich insights into the temporal and spatial distribution of topics, sentiment, and substantive themes within global news coverage focusing on the latest emerging technology --generative AI. We collected a comprehensive dataset of news articles (January 2018 to November 2023, \textit{N} = 24,827). For topic modeling, we employed the BERTopic technique and combined it with qualitative coding to identify semantic themes. Subsequently, sentiment analysis was conducted using the RoBERTa-base model.
Analysis of temporal patterns in the data reveals notable variability in coverage across key topics--business, corporate technological development, regulation and security, and education--with spikes in articles coinciding with major AI developments and policy discussions. Sentiment analysis shows a predominantly neutral to positive media stance, with the business-related articles exhibiting more positive sentiment, while regulation and security articles receive a reserved, neutral to negative sentiment. Our study offers a valuable framework to investigate global news discourse and evaluate news attitudes and themes related to emerging technologies.
\end{abstract}

\section{Introduction}

Media plays an important role in disseminating information to the public and shaping their perceptions of emerging technologies \cite{brossard2013new}. News reporting on artificial intelligence (AI)-related content can help foster essential technology literacy among the general public \cite{nguyen2022news}. Studying generative AI is crucial, not only for its advanced capabilities and applications beyond traditional machine learning technologies but also due to the emerging challenges and societal implications, which require specialized understanding and careful scrutiny. Despite the significant potential of Generative AI to transform various of industries and aspects of public life\cite{Marr.2023}, public understanding of this emerging technology remains nascent. By analyzing news media coverage of generative AI, researchers can understand how the public is exposed to media interpretations of data and events related to this innovative technology, and how this may form public opinions or judgments about generative AI.

In the realm of AI-related news, prior research has investigated the broader context of AI news, often emphasizing Western mainstream media outlets. However, the investigation into the reporting of Generative AI news remains understudied, particularly from a global perspective \cite{sun2020newspaper,ChuanTsaiCho2019}. As a result, the global media's response to the rapidly evolving generative AI industry remains unclear. Our work aims to bridge these gaps by analyzing a large corpus of global AI-related news coverage. This study also seeks to illuminate the general public’s exposure, sentiment, and perceptions toward the latest advancements in innovative technology. \par

Our analysis is based on a large-scale dataset consisting of English-language news articles about large language models (LLMs) published by a total of 703 US national, US local, and international news outlets. This dataset includes over 24,827 articles collected over a period of 70 months, from January 2018 to November 2023, allowing for a comprehensive analysis of the evolution and impact of LLMs since their conception. To the best of our knowledge, our data represents the most comprehensive and current news resource available, capturing the global discourse surrounding LLMs. In this paper, we address the following research questions: \par

\textbf{RQ1}: \textit{After the emergence of Generative AI, what topics do news articles about it focus on, and how does the quantity of news articles on these topics vary temporally and spatially?}

\textbf{RQ2}: \textit{How does the sentiment of news articles about Generative AI vary across topics and the different categories of news outlets?} 

\textbf{RQ3}: \textit{What are the most popular topics covered by Generative AI news with positive sentiment?} \par

We found that after the introduction of ChatGPT in late 2022, news coverage of generative AI has been marked by temporal and spatial variabilities in the number of news articles. Those trends coincide with major events regarding technological development and specific interests. Our sentiment analysis reveals that generative AI is predominantly portrayed in a neutral to positive light, echoing the optimistic tone towards emerging technologies consistently documented in prior research \cite{GarveyandMaskal2020, Fast_Horvitz_2017}. However, in the context of generative AI, our analysis uncovers a more reserved sentiment in reports focusing on regulation and security reporting. 

 
 \par

This work makes the following contributions: 

\begin{enumerate}
  \item First, we gathered a diverse range of global news articles spanning various types of news outlets, countries, and time \footnote{Our dataset is available upon request.}. This collection extends from the initial release of BERT in 2018 to November 2023, allowing a comprehensive understanding of the global news coverage and perspectives on the rapidly evolving generative AI technologies since the inception of LLMs. This dataset can also be used to continue exploring the global news discourse around generative AI and other emerging technologies.

 \item Second, we constructed an extensive codebook for the topics of generative AI news articles, which builds on both quantitative topic modeling and qualitative manual coding results. This codebook encompasses applications of generative AI across sectors, such as business and education, as well as responses to generative AI like regulation and security. Our methodology and the developed codebook offer a valuable framework for future researchers to capture and assess the news coverage of future emerging technologies.
 
 \item Third, we utilized two approaches, sentiment and semantic analysis, to further capture the attitudes and themes present in global news coverage of generative AI technologies. Aligning with the overall optimistic views of emerging technologies in early studies, our analysis also reveals a reserved tone in coverage related to regulatory and security aspects of generative AI. This dichotomy highlights the complexity of global media coverage regarding the integration of generative AI into various industries and the social fabric, emphasizing the nuanced nature of its reception across diverse contexts. 
 
 \end{enumerate}

\par

\section{Related Works}

While prior research has not explicitly focused on analyzing news coverage of Generative AI, several studies have delved into the broader context of AI news coverage and provided nuanced insights into the specific themes and sentiment analysis over time. 

One study analyzed five major American newspapers (i.e., \textit{USA Today}, \textit{The New York Times (NYT)}, \textit{Los Angeles Times}, \textit{New York Post}, and \textit{Washington Post}) from 2009 to 2018 and found that business and technology were the predominant subjects in AI news coverage \cite{ChuanTsaiCho2019}. Another study analyzed the \textit{New York Times}, \textit{Washington Post}, \textit{the Guardian}, and \textit{USA Today} from 1977 to 2019, identified fourteen major topics, including research and education, media products, health care, jobs, economy, and others \cite{sun2020newspaper}. 

Previous research found mixed sentiment analysis evidence for AI news reporting. An automated content analysis reveals the rapid emergence of AI's ubiquity in the mid-2010s and demonstrates a growing critical tone in news discourse over time among \textit{The NYT}, \textit{The Guardian}, \textit{Wired}, and \textit{Gizmodo} \cite{nguyen2022news}. However, others found that the majority AI news reporting was positive over six decades (1956 to 2018) among The \textit{NYT}, \textit{Associated Press}, \textit{The International Herald Tribune}, \textit{Reuters}, \textit{CNBC}, \textit{International NYT}, and \textit{Internet Video Archive}) \cite{GarveyandMaskal2020}. Similarly, the analysis of a 30-year news reporting from the \textit{New York Times} also revealed overall consistently optimistic tones for AI news coverage \cite{Fast_Horvitz_2017}. 

While discussions about the benefits of AI were more frequent than its risks, the discussions on AI risks were generally more specific \cite{ChuanTsaiCho2019}. News reporting of AI showed growing concerns about loss of control, ethical issues, and negative impacts on work in recent years, despite of increasing hopes for AI in healthcare and education \cite{Fast_Horvitz_2017}. Leading English-speaking global media outlets reported concerns include privacy invasion, data bias, cybersecurity, and information disorder, underscoring the importance of interventions to clarify the detrimental impacts of datafication and automation on citizens \cite{nguyen2023news}. Journalists portrayed AI as sophisticated, powerful, and value-laden, but the perspectives of ordinary citizens were notably absent in media discourses \cite{sun2020newspaper}.

\section{Data}

\subsection{Data Preparation}

We collected news data from ProQuest TDM \cite{tdmstudio}, a platform encompassing multiple databases across disciplines, including newspapers, magazines, dissertations, and other primary sources. To collect relevant news, we conducted a search based on the following conditions as presented in Table \ref{table: search conditions}: date, search terms, and ProQuest newsstream databases. Each ProQuest newsstream database covers an array of news outlets. For example, North Central Newsstream databases curate news articles published by state-level news outlets in Colorado, Iowa, Kansas, Minnesota, Missouri, Montana, North Dakota, Nebraska, South Dakota, and Wyoming \cite{tdmstudio}.

\begin{table*}[t]
\begin{tabular}{l|p{0.81\textwidth}}
    \hline \hline
    \textbf{Search Category} & \textbf{Search Conditions} \\
    \hline
    Date & 2018-01-01 to 2023-11-18  \\
    Search terms & Large language model, LLM, ChatGPT, BERT, GPT, PaLM, LLaMA \\
    Newsstream & African Newsstream, Asian Newsstream, Australia Newsstream, New Zealand Newsstream, Canadian Newsstream, European Newsstream, Latin American Newsstream, Middle East Newsstream, U.S. Hispanic Newsstream, U.S. Midwest Newsstream, U.S. North Central Newsstream, U.S. Northeast Newsstream, U.S. South Central Newsstream, U.S. Southeast Newsstream, U.S. West Newsstream \\
    \hline \hline
\end{tabular}
\caption{Search conditions to collect related news articles from ProQuest.}
\label{table: search conditions}
\end{table*}


We established our search time frame from January 2018 to November 2023, aligning with the initial release of the widely acclaimed Bidirectional Encoder Representations from Transformers (BERT) in 2018. Our focus for search terms included ``large language model," along with popular LLMs such as BERT, OpenAI's GPT, Google's PaLM, and Meta's LlaMa. In addition, we included a list of popular ProQuest newsstream databases worldwide \cite{tdmnewsstream}. Collectively, these criteria resulted in a total of 38,199 news articles.

We refined the acquired dataset in several ways. First, after detecting the language of news articles, we included only articles that were written in English for later analysis, which is the predominant language in our dataset. Second, we removed duplicated news articles based on their title, content, and publisher. 
After reading 100 randomly selected copies of news articles, we found that the content of many articles was not related to generative AI but was included in our dataset. This is because common names of popular LLMs (e.g., ``PaLM'') overlap with regular expressions (e.g., ``palm''). Thus, we implemented a third filter requiring the mention of popular LLMs alongside their respective company names. For instance, ``PaLM" needed to be paired with ``Google''. Implementing these conditions resulted in a final dataset comprising 24,827 news articles for subsequent analysis.

Prior to topic modeling, we further cleaned the text. This process included removing short URLs, digits, emojis, and punctuation from news. Additionally, non-informative stop-words like ``the,'' ``is," and ``and'' were eliminated from the text. Subsequently, we tokenized each piece of news into individual words and characters and then lemmatized to their base or stemming forms. The resulting dataset contains 24,827 English news articles, with their full text, title, author, publication date, publication country, and publication address.

\subsection{Data Profiling}

Our dataset, sourced from the ProQuest Newsstream database, comprises 24,827 English-language news articles from a diverse array of publications. To give an overview of the dataset, we categorized these news outlets into three groups, based on the categorization used by ProQuest Newsstream databases, such as \cite{proquestmajor}, and corroborated by existing literature, such as \cite{shearer2021broad}. The groups are: US national news outlets, US local and specialized news outlets, and international news outlets. The categorization details are provided in Table \ref{table: news outlets}.

\begin{table*}[t]
\begin{tabular}{p{0.15\textwidth}|p{0.18\textwidth}|c|p{0.52\textwidth}}
    \hline \hline
    \textbf{News outlet} & \textbf{Pub title} & \textbf{Count} & \textbf{Sampled news title} \\
    \hline
    US national news & Wall Street Journal & 607 & ``OpenAI to Offer ChatGPT Subscription Plan for \$20 a Month;'' ``How Worried Should We Be About AI’s Threat to Humanity? Even Tech Leaders Can’t Agree'' \\
    US national news & USA Today & 176 & ``Why Elon Musk wants to build ChatGPT competitor: AI chatbots are too ‘woke’;'' ``Sears pioneered the modern prefab house in the early 20th century: Look back at ‘kit homes’'' \\
    US national news & The New York Times & 128 & ``Microsoft Says New A.I. Shows Signs of Human Reasoning;'' ``Google Tests an A.I. Assistant That Offers Life Advice'' \\
    US local and specialized news & PR Newswire & 1921 & ``Pioneering Real-World AI: Wecover Platforms Brings Generative AI Experience to MBA Students at Georgia State University;'' ``Treehouse Adopts AI to Help Students Prepare for the Next Great Technology Wave'' \\
    US local and specialized news & Business Wire & 1155 & ``Folloze AI, Powered by ChatGPT, adds Critical Layer of Buyer Engagement Insights to Drive Increased Revenue;'' ``Flatiron School Launches New Artificial Intelligence Training Programs'' \\
    US local and specialized news & Barron's & 431 & `` \$90 Billion Valuation for OpenAI? Tech's New Star Is Red Hot;'' ``AI to Pick Stocks? JPMorgan's Move Hints at Banks Following Big Tech's Lead'' \\
    International news & The Times of India & 753 & ``Google to launch new chatbots for advertisers and YouTube content creators;'' ``Samsung bans use of ChatGPT and other AI tools for employees'' \\
    International news & The Guardian & 369 & ``Monday briefing: What the AI boom really means for your job (and mine);'' ``Everything you wanted to know about AI but were afraid to ask'' \\
    International news & South China Morning Post & 275 & ``Baidu's ChatGPT alternative gets positive reviews for handling of Chinese translations as search giant's stock jumps;'' ``Alibaba tests ChatGPT rival as Chinese tech giants like Baidu race to build country's best AI chatbot''  \\
    \hline \hline
\end{tabular}
\caption{Representative examples of categorization of news outlets.}
\label{table: news outlets}
\end{table*}

Within the US national news outlets category, our dataset includes publications like the Wall Street Journal, which alone contributes 607 articles and accounts for 60\% of the articles in this category. Other major US national news outlets in our dataset include USA Today (228, 22\%) and The New York Times (128, 13\%). These outlets are generally recognized for their national presence and have a significant impact on shaping public opinion and national discourse. 

Popular US local and specialized news outlets and publishers in our dataset include: PR Newswire (1921, 24\%), Business Wire (1155, 15\%), Targeted News Service (1124, 14\%), University Wire (723, 10\%), Politico (348, 4\%), and Boston Globe (111 2\%). Publishers like PR Newswire and Business Wire usually disseminate press releases and corporate news, while specialized news outlets like Politico focus on political journalism and in-depth coverage of Washington D.C. and policy, and local news outlets like Boston Globe cater to specific local or regional audiences with news that resonates with their immediate geographical and cultural context. 

Within the international news outlets category, top outlets in our dataset are: India's The Times of India (753, 5\%), Indian Express (645, 4\%), Mint (4\%), IANS English (471, 3\%), and Financial Express (448, 3\%); UK's Telegraph.co.uk (393, 2\%) and The Guardian (369, 2\%); and Thailand's Asia News Monitor (286, 2\%). These outlets represent a rich source of international English news coverage and contribute to the geographical diversity of our dataset, especially in Asia.


\section{Methodology}

\subsection{BERT Topic Modeling}

Topic modeling allows for the identification of semantic themes in a large volume of textual data \cite{vayansky2020review}. To mine the themes from the collected news data, we applied the BERTopic technique, which involves using BERT word embedding to extract semantically relevant sentence embeddings from documents. We chose BERTopic over Latent Dirichlet Allocation (LDA) for topic modeling due to its distinct advantage in understanding the semantic meanings of words with contextualized representation \cite{reimers2019sentence}. Prior research has also demonstrated that BERTopic performs better than LDA and Top2Vec in identifying topics extracted from online posts \cite{egger2022topic}. 

Due to the high dimensionality of the vectors generated by the BERT embedding from news text that presents a challenge for machine processing, we utilized a dimensionality reduction technique called Uniform Manifold Approximation and Projection (UMAP) proposed by McInnes et al. (2018) \cite{mcinnes2018umap}. The UMAP method can help to mitigate high dimensionality issues while retaining the local and global structure of the dataset \cite{mcinnes2018umap}. 

\begin{figure}[t]
    \centering
    \includegraphics[width=1\columnwidth]{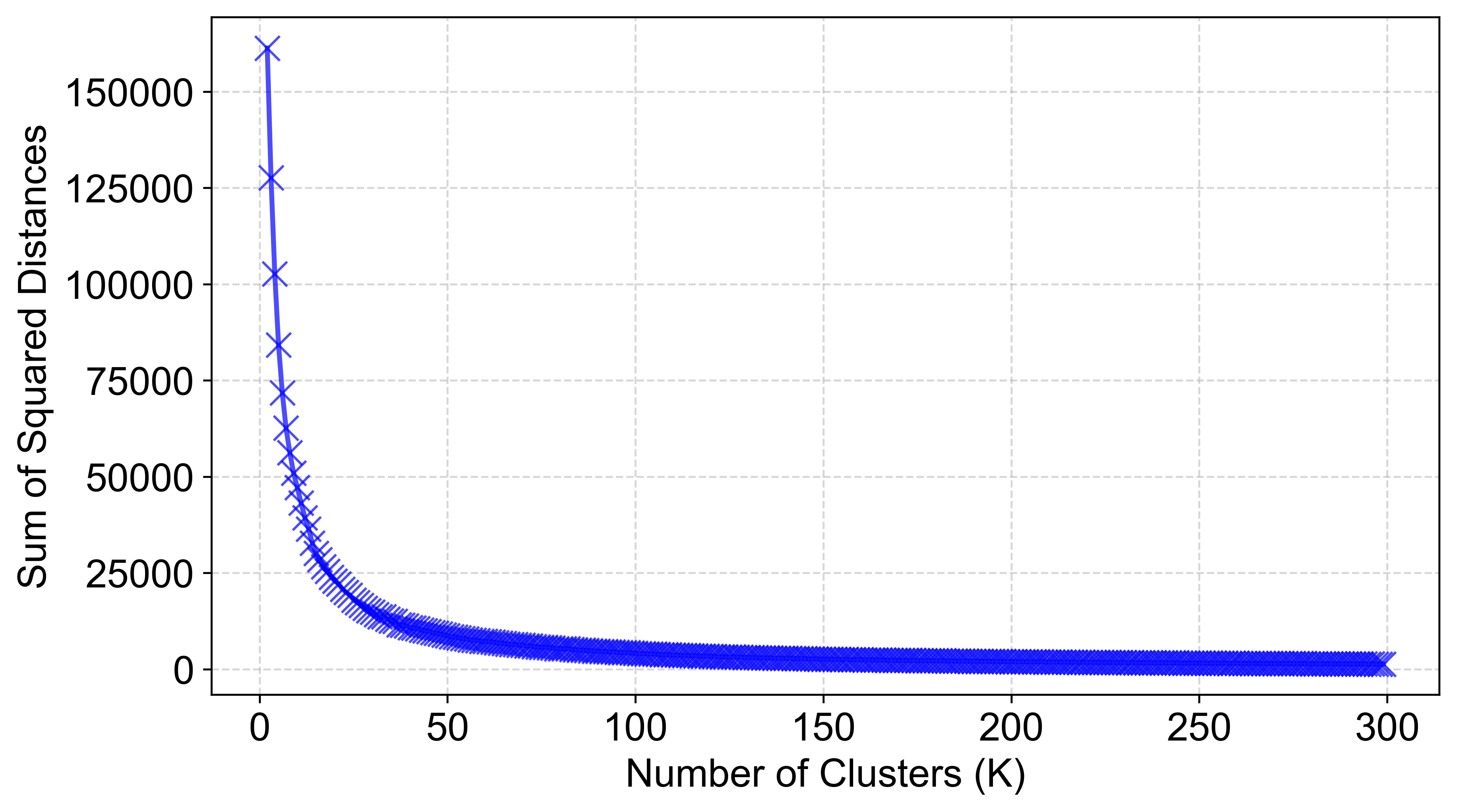}
    \caption{Result of the Elbow method to determine the optimal K clusters in BERTopic modeling.}
    \label{fig:elbow}
\end{figure}

Subsequently, we used the elbow method in conjunction with K-means to determine the optimal number of clusters for topic modeling. The elbow method is a graphical method that allows us to find the best K clusters by assessing the Within-Cluster Sum of Squares (WCSS) — the summation of squared distances between cluster points and their centroids. We implemented the experimentation using numbers ranging from 2 to 300 clusters, as depicted in Figure~\ref{fig:elbow}. This figure shows a significant reduction in WCSS, around 50 clusters. For a more refined clustering result, we chose to use 100 clusters to manage our collected news data. This handling enhances granularity and precision while allowing the research team to use a bottom-up approach (i.e., manually annotating each of the 100 clusters into categories) to validate these clusters. With the determined optimal $K=100$, we applied K-Means clustering \cite{buitinck2013api} to group articles in the dataset into 100 clusters. Specifically, we applied the model to the main text of news articles, given that it holds richer information compared to news titles. 

The final stage of our topic modeling process involves the representation of topics. We used a count vectorizer technique called the class-based Term Frequency-Inverse Document Frequency (c-TF-IDF) within the Scikit-learn Python package \cite{grootendorst2022bertopic} to tokenize the topics. This method can help to extract the topical keywords and representative documents from each cluster.

\subsection{Qualitative Coding}

Using BERTopic to cluster news articles prompted a subsequent need to categorize the topics of the clusters of news articles. To do so, we first employed an iterative approach that combined both top-down and bottom-up processes to identify the potential topics. For consistency, we used the term ``cluster'' to refer to the clusters returned by BERTopic and the term ``topic'' to refer to the manually identified topic information from each cluster. In the top-down phase, we referenced relevant papers, such as the work by \cite{sun2020newspaper}, which identified 14 key topics in news articles related to emerging AI technologies. These encompassed education, healthcare, job market, life science, business, risk assessment, art and creation, regulation and policy, gaming, software, transportation, robotics, and algorithms. Drawing insights from prior research enabled us to collect potential general topics the news articles cover regarding emerging AI technology. For the bottom-up process, we reviewed a random sample of 100 collected articles and manually assigned topics to each article based on our understanding and expertise on the subject. This bottom-up process ensured that the collected topics aligned with our specific context and contributed to the completeness of our compiled list.

Upon collecting candidate topics, four authors of the research team engaged in a brainstorming session to discuss and refine the identified topics. Following three iterations, we finalized the topics into a set of 10 topics, including education, business, labor and work, corporate technological development, regulation and security, content creation, healthcare, politics, finance, and a ``miscellaneous" category covering areas like engineering applications and robot development. We removed articles that are tangentially related to generative AI in this process, which amounts to about 10\% of the entire dataset.

Next, we employed a ``paired-coding" method to label each cluster output by BERTopic. That is, each cluster was guaranteed to be labeled by two authors. During the labeling process, each author first examined the representative word list generated through the c-TF-IDF method in the BERTopic model. Then, we reviewed a sample of news articles within each cluster. Next, each author independently labeled the cluster and assigned it to one of the ten predetermined topics. During this labeling process, we also observed a few clusters that were unrelated to generative AI settings, which were excluded from subsequent analysis.

After each author completed the labeling, we employed Krippendorff’s $\alpha$ \cite{krippendorff2018content}, an inter-coder reliability index, to quantify the level of agreement between independent annotators. In our experiment, the calculated Krippendorff’s $\alpha$ is 0.88. As suggested, a Krippendorff’s $\alpha$ above 0.8 is generally indicative of good agreement \cite{krippendorff2018content}. For clusters that received different labels, we conducted an additional brainstorming session. Through the discussion, all authors reached a consensus on the annotation for each of these clusters. The results of the topic annotation are listed in the table below.

\begin{table*}[t]
\begin{tabular}{>{\raggedright\arraybackslash}p{0.15\textwidth}|>{\raggedright\arraybackslash}p{0.25\textwidth}|p{0.07\textwidth}|p{0.07\textwidth}|>{\raggedright\arraybackslash}p{0.33\textwidth}}
    \hline \hline
    \textbf{Topic} & \textbf{Explanation}  & \textbf{\#Clusters} 
    & \textbf{\#Articles} & \textbf{Representative clusters (keywords)} \\
    \hline
    Corporate technological development & Development and innovation of generative AI led by corporate companies  & 17 
    & 4798 & chatgpt-language-model-user, bard-google-search-pichai, microsoft-copilot-openai-bing, user-image-content \\
    Regulation and security & Risks, ethical issues, and policies of AI, including AI regulation in global context & 15 
    & 4031 & ai-congress-agency-government, attack-cybersecurity-cyber-phishing, security-cybersecurity-data-threat, eu-european-act-parliament \\
    Business & Impact on customer service and user experience, product management, marketing, and retail  & 9 
    & 3177 & data-cloud-customer-enterprise, customer-business-solution-data, generative-ai-business-customer, brand-customer-retailer-consumer \\
    Education & Generative AI's applications in the school context, and its impact on students, teachers, and professors  & 8 
    & 2710 & student-school-teacher-education, student-college-professor-university, science-research-university-student, text-chatgpt-write-paper \\
    Labor and work & Generative AI's impact on labor force and job market & 6 
    & 1925 & job-worker-ai-work, job-hr-employee-skill, news-journalist-journalism-medium \\
    Content creation & The use of generative AI in creative fields, including writing, music, and art & 5 
    & 1117 & writer-film-actor-strike, music-song-artist-elvis, art-image-artist-create \\
    Finance & Generative AI's impact on investment, stock, cryptocurrency, or company revenue & 4 
    & 1049 & investor-stock-investment-fund, financial-company-investment-investor, rypto-cryptocurrency-worldcoin-blockchain \\
    Healthcare & Generative AI's applications in analyzing biomedical information and healthcare  & 6 
    & 1043 & health-patient-healthcare-care, medical-patient-doctor-health, drug-protein-cell-disease \\
    Politics & Generative AI's impact on US politics and elections, and global politics of AI & 5 
    & 1008 & trump-republican-ramaswamy-debate, labour-political-conservative, japan-kishida-minister-japanese \\
    Miscellaneous & Generative AI's applications in other areas such as food and environment & 7 
    & 1371 & energy-water-carbon-climate, food-restaurant-recipe-eat \\ 
    \hline \hline
\end{tabular}
\caption{Topic identification and examples of representative clusters within each topic.}
\label{table: topic_modeling}
\end{table*}

\subsection{Sentiment analysis}

With each news article categorized into a topic, we employed sentiment analysis to identify whether the new article carries a positive, neutral, or negative sentiment. This allowed us to further make the comparative analysis from both temporal and spatial perspectives regarding the sentiment. We used the RoBERTa-base model \cite{barbieri2020tweeteval, loureiro2022timelms} to implement the sentiment analysis. 

RoBERTa is a robustly optimized Bidirectional Encoder Representations (BERT) pre-training approach based on BERT embedding. BERT embedding is built on transformer's architecture and attention mechanisms to create contextualized representations from text. The RoBERTa-based sentiment model was fine-tuned for sentiment analysis using the TweetEval benchmark. \cite{barbieri2020tweeteval} also enhanced the model by integrating a dense layer to reduce the dimensions in the last layer to match the number of classifications in the sentiment task. This model has been demonstrated to outperform FastText and Support Vector Machine (SVM)-based models employing n-gram features. Specifically, it achieved an accuracy of about 72\% on the TweetEval testing dataset, as compared to the 63\% accuracy achieved by SVM and FastText \cite{barbieri2020tweeteval}.

Loureiro et al. (2022) further enhanced the model using an expanded training corpus of social media postings, consisting of 123.86 million tweets collected until the end of 2021, a notable increase from its predecessor’s 58 million tweets. This update resulted in an improved performance of sentiment analysis, demonstrating an accuracy of 73.7\% on the TweetEval dataset \cite{loureiro2022timelms}.

\section{Findings}

Manually coding the news articles by a variety of themes gave us the following distribution of news: there are 4798 articles (19\%) on corporate technological development, 4031 articles (16\%) on regulation and security, 3177 articles (13\%) on business, 2710 articles (10\%) on education, 1925 articles (10\%) on labor and work, 1117 articles (8\%) on content creation, 1049 articles (6\%) on finance, 1043 articles (4\%) on healthcare, and 1008 articles (4\%) on politics. About \%6 articles cover other miscellaneous topics. Table~\ref{table: topic_modeling} presents the codebook used in qualitative coding and the relationship between topics of articles and BERTopic clusters in more detail, with examples of representative BERTopic clusters for each topic.

\subsection{RQ1: When were news articles of different topics published and where?}

\subsubsection{Temporal distribution of articles by topic} 

\begin{figure}[t]
    \centering
    \includegraphics[width=1\columnwidth]{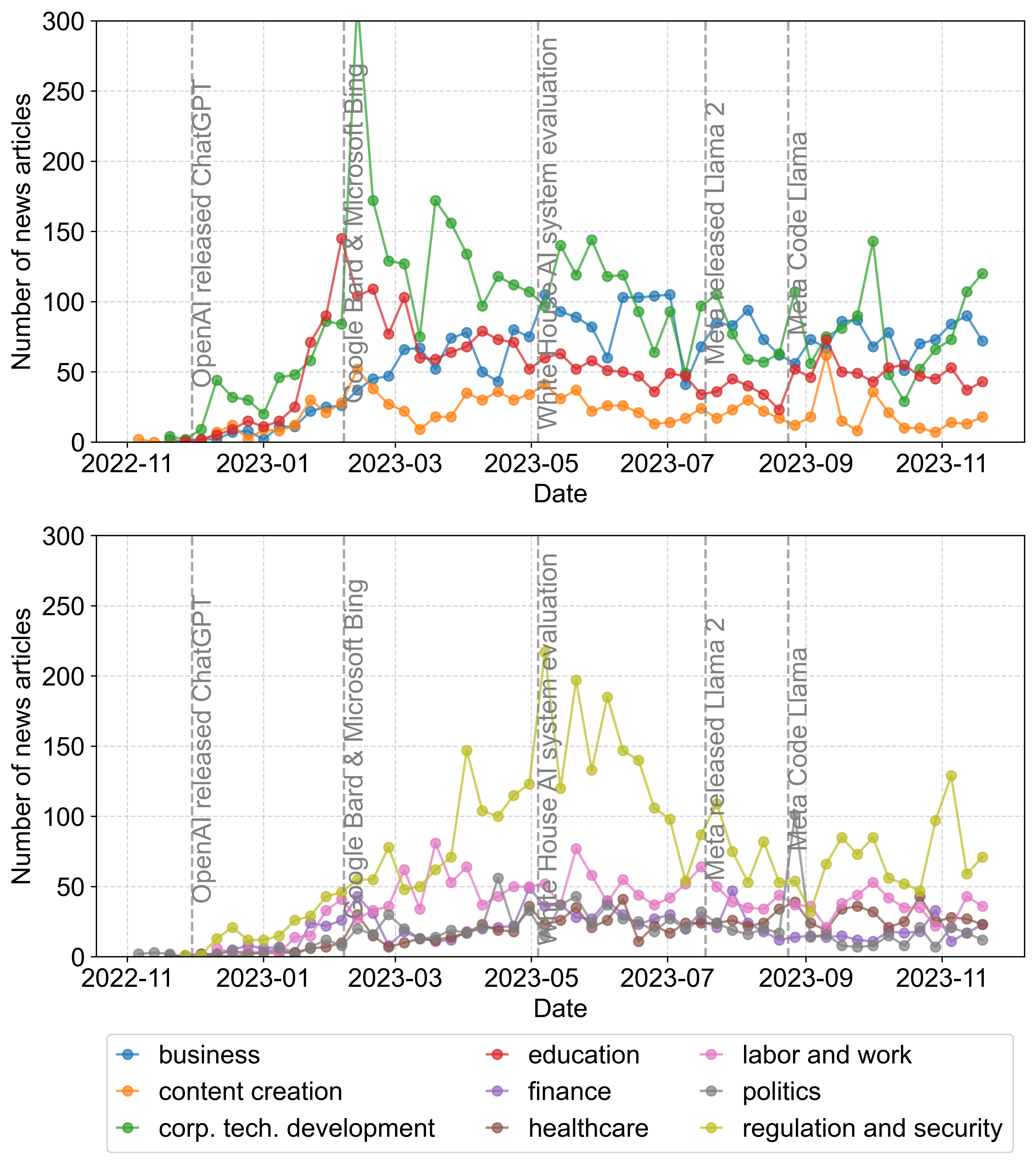}
    \caption{Temporal trends of news articles across topics.}
    \label{fig:temporal}
\end{figure}

Based on topic modeling and qualitative coding results, we analyzed the temporal distribution of article topics, centering on articles published following the introduction of ChatGPT in November 2022. Figure \ref{fig:temporal} shows a time series of the number of articles across different topics. The dominant topics throughout this period—business, corporate technological development, regulation and security, and education—underscore the impact of generative AI across these topics. Over the span of a year, from November 2022 to November 2023, we observed notable variability in the volume of articles across topics. The moments of high news coverage may be related to key developments or product releases of generative AI, the shifting focus areas in news coverage, as well as the news' responsiveness to specific events in the course of new technology development. 

For example, the topic of corporate technological development shows a significant spike in articles in the weeks of February 12, 2023, and February 19, 2023. This could indicate a surge of interest or events related to the introduction of Bard by Google on February 6 and Microsoft's launch of an updated version of its Bing search engine on February 7. In another example, the topic regulation and security exhibits a surge in the week of May 7, 2023, which might be related to the White House's meeting with CEOs of large technology companies on advancing responsible AI innovation on May 4. The variability and volume of coverage for each topic would be indicative of their relevance and newsworthiness.

\subsubsection{Spatial distribution of articles by topic}

\begin{figure}[t]
    \centering
    \includegraphics[width=1\columnwidth]{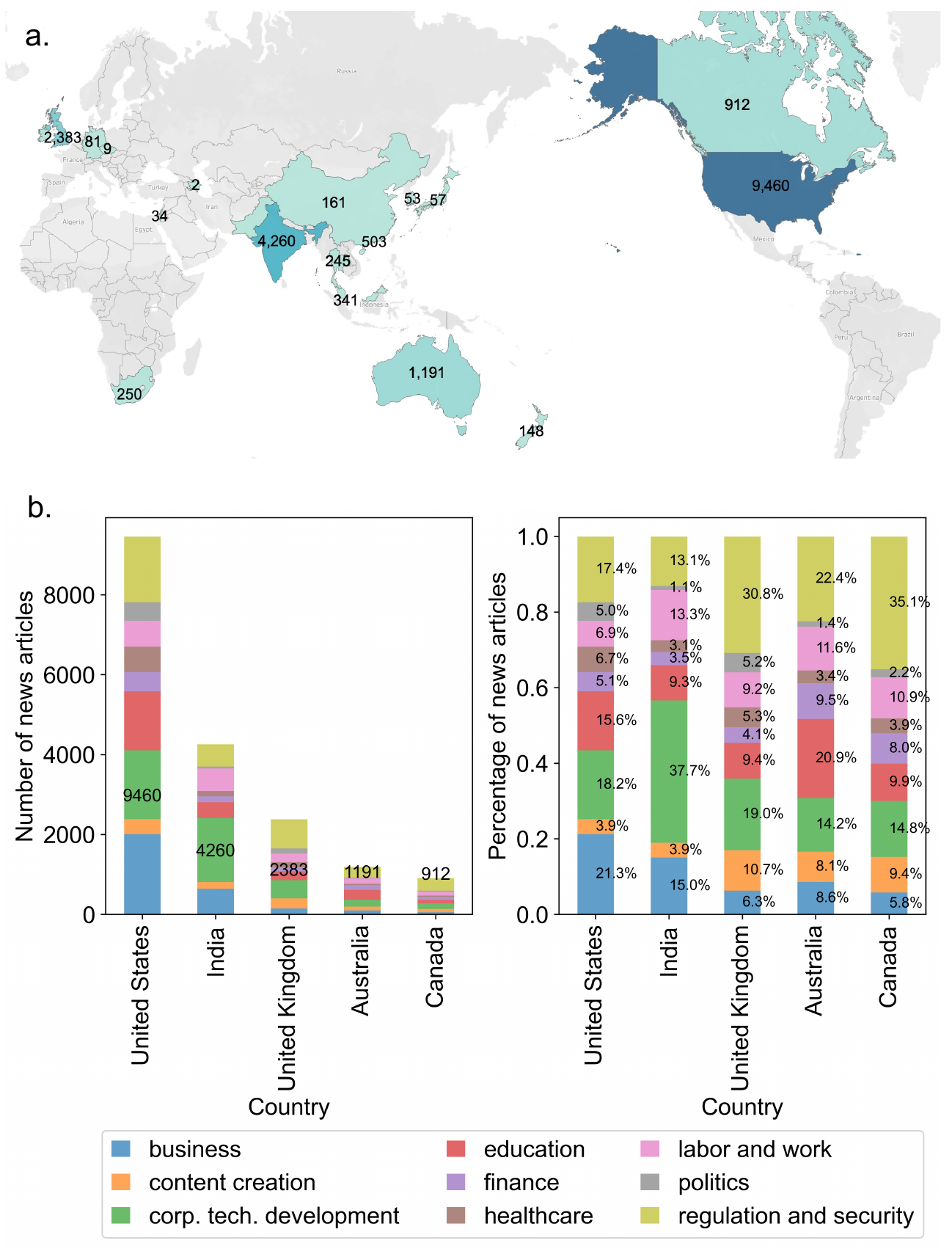}
    \caption{Collected English news articles (a) Spatial distribution across countries and regions. (b) Count and percentage of news articles of top five countries across topics.}
    \label{fig:country}
\end{figure}

In addition, we found distinct patterns reflective of regional focus and interests by examining the spatial distribution of news articles by topic. For this analysis, we focused on the top five countries by article count (US, India, UK, Australia, and Canada), as illustrated in Figure \ref{fig:country}.

In the US, the topic of business constitutes the largest segment of articles (2013 articles, 21\% of all articles in this country), followed closely by corporate technological development (1725, 18\%) and education (1477, 16\%). The prominence of news coverage on those topics may reflect the country's news attention on business-related technological implications, technological advancements by corporate companies, and the widespread impact of generative AI on educational frameworks.
India's news distribution shows an overwhelming majority of articles concentrated on corporate technological development (1605, 38\%). This focus could be indicative of the country's burgeoning tech industry and the rapid development of technology that drives media attention.
The UK exhibits a more balanced distribution, with a substantial portion of articles dedicated to regulation and security (734, 31\%), followed by corporate technological development (452, 19\%). This could suggest a heightened awareness and engagement with generative AI in terms of governance and the implications for security frameworks within the region.
In Australia, similar to the UK, a significant share of coverage is directed at regulation and security (267, 22\%), suggesting a national focus on generative AI's governance and ethical implications. This is closely followed by education (249, 21\%), which might reflect the country’s prioritization of educational initiatives in generative AI.
Canada's news articles display a strong emphasis on regulation and security (320, 35\%), which is the most prominent topic, overshadowing corporate technological development (135, 15\%) and other topics. This might suggest Canada's proactive stance in addressing the complexities of AI governance and the associated security challenges.

Comparing the news coverage across countries reveals distinct national interests and priorities in generative AI. For instance, corporate technological development and business are leading topics in the US and India, which could reflect their positions as major hubs of technological innovation. In contrast, the UK, Australia, and Canada show a heightened interest in regulation and security, pointing to a more cautious and governance-oriented approach. Australia’s focus on education suggests an investment in developing human capital to navigate and leverage AI advancements.

\subsection{RQ2: What is the sentiment of articles?}

We utilized a RoBERTa-based model to determine the sentiment of each article, which categorized them into negative, neutral, and positive sentiments. The overall discourse in the news articles on generative AI leans towards a neutral and positive portrayal, with 66\% neutral and 28\% positive sentiment articles.
The tendency towards positive sentiment across diverse topics may reflect an overall optimistic or progressive narrative in media reporting or a trend in the media towards focusing on positive aspects or developments within these areas, which has been documented by prior scholarship \cite{GarveyandMaskal2020}. In this section, we delve deeper into the news articles by looking into the topic, type of news outlet, and content to capture the nuances behind the predominance of neutral and positive sentiments.

\subsubsection{Temporal sentiment of articles by topic}

For each of the top four topics (business, corporate technological development, regulation and security, and education), we computed weekly sentiment scores for articles within the topic for the time frame from November 2022 to November 2023. The sentiment scores were calculated using a scale where positive sentiment was assigned a value of 1, negative sentiment a value of -1, and neutral sentiment a value of 0. These values were then averaged on a weekly basis to produce a score ranging from -1 to 1 to normalize for comparison.

\begin{figure}[t]
    \centering
    \includegraphics[width=1\columnwidth]{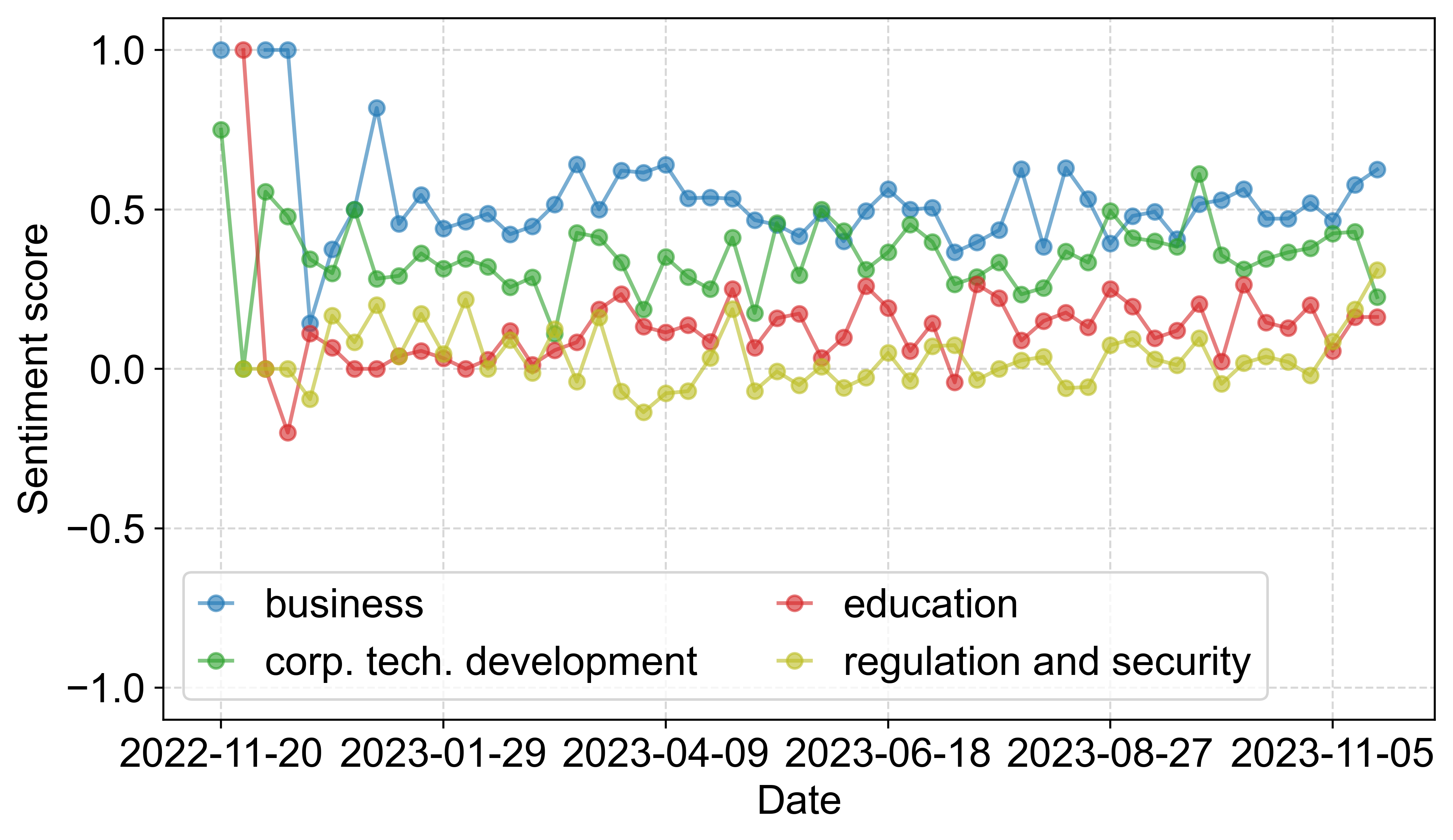}
    \caption{Weekly sentiment scores from November 2022 to November 2023, for the top four topics: corporate technological development, regulation and security, business, and education.}
    \label{fig:sentiment}
\end{figure}

Figure \ref{fig:sentiment} presents a weekly sentiment analysis trend of articles across four topics: corporate technological development, regulation and security, business, and education. There are observable differences in sentiment across the four topics. Business and corporate technological development generally show more positive sentiment scores compared to the other topics, indicating news articles on these topics are often reported with a positive tone. The topic of education also trends positively, though with some variation, suggesting a generally favorable portrayal in the news. Regulation and security articles exhibit a markedly lower sentiment, often hovering around the neutral to negative areas. This trend may reflect the inherently critical nature of news related to regulatory measures and security issues, which often involve reporting on conflicts, debates, and challenges within the realm of AI governance. As such, the sentiment of news articles may influence public perception and discourse around these aspects of generative AI.

\subsubsection{Sentiment of articles by news outlet}

Sentiment towards generative AI varies significantly across different types of news outlets. For each of the three news outlet categories, we obtain a list of sentiment scores using the same scale as before, where positive sentiment was assigned a value of 1, negative sentiment a value of -1, and neutral sentiment a value of 0. 

We performed the non-parametric Kruskal-Wallis H-test test on sentiment scores of three groups of news outlets: US national, US local and specialized, and international news outlets. With a test statistic of 141.12 and a significant p-value ($\leq$ 0.001), there is a statistically significant difference in sentiment scores among these groups. This result is further substantiated by the pairwise comparisons of Dunn's test in Table~\ref{table:anova_outlets} that accounted for multiple comparisons.
Specifically, international news outlets are more neutral compared to US national news outlets, while US local and specialized news outlets show a tendency towards more positive sentiment. This may imply that US local and specialized outlets are experiencing more positive impacts or are more optimistic about the potential of generative AI. Additionally, the prevalence of positive sentiment in US local and specialized news might be related to their sourcing strategies, which could include a higher proportion of press releases from news wire services. Conversely, US national news outlets may adopt a more cautious or critical stance, reflecting their broader audience and perhaps a responsibility to present a more balanced view of the developments in generative AI. 

\begin{table}[t]
\centering
\begin{tabular}{c|c|c|c}
    \hline \hline
    \textbf{Type 1} & \textbf{Type 2} & 
    \textbf{Mean Diff} & 
    \textbf{Adj. p-value} \\
    \hline
    International & US national & -0.069 & 0.0*** \\
    International & US local & 0.0813 & 0.0*** \\
    US national & US local & 0.1503 & 0.0*** \\
    \hline 
    \multicolumn{4}{l}{*$p \leq$ 0.05, **$p \leq$ 0.01, ***$p \leq$ 0.001} \\
    \hline \hline
\end{tabular}
\caption{Dunn's test for multiple pairwise comparison of news outlet types, as well as the mean difference in sentiment scores between news outlet types.}
\label{table:anova_outlets}
\end{table}

\subsection{RQ3: What are the articles about?}

In this section, we delved into the substantive themes within generative AI news discourse. Building on the spatiotemporal and sentiment analysis conducted in RQ1 and RQ2, which indicated a predominantly positive media perspective across various topics, RQ3 seeks to elucidate the content that characterizes these discussions. Considering the scope of the paper, we focused on the four primary topics in news coverage of generative AI: business, corporate technological development, regulation and security, and education.

\begin{figure*}[!t]
    \centering
    \includegraphics[scale=0.35]{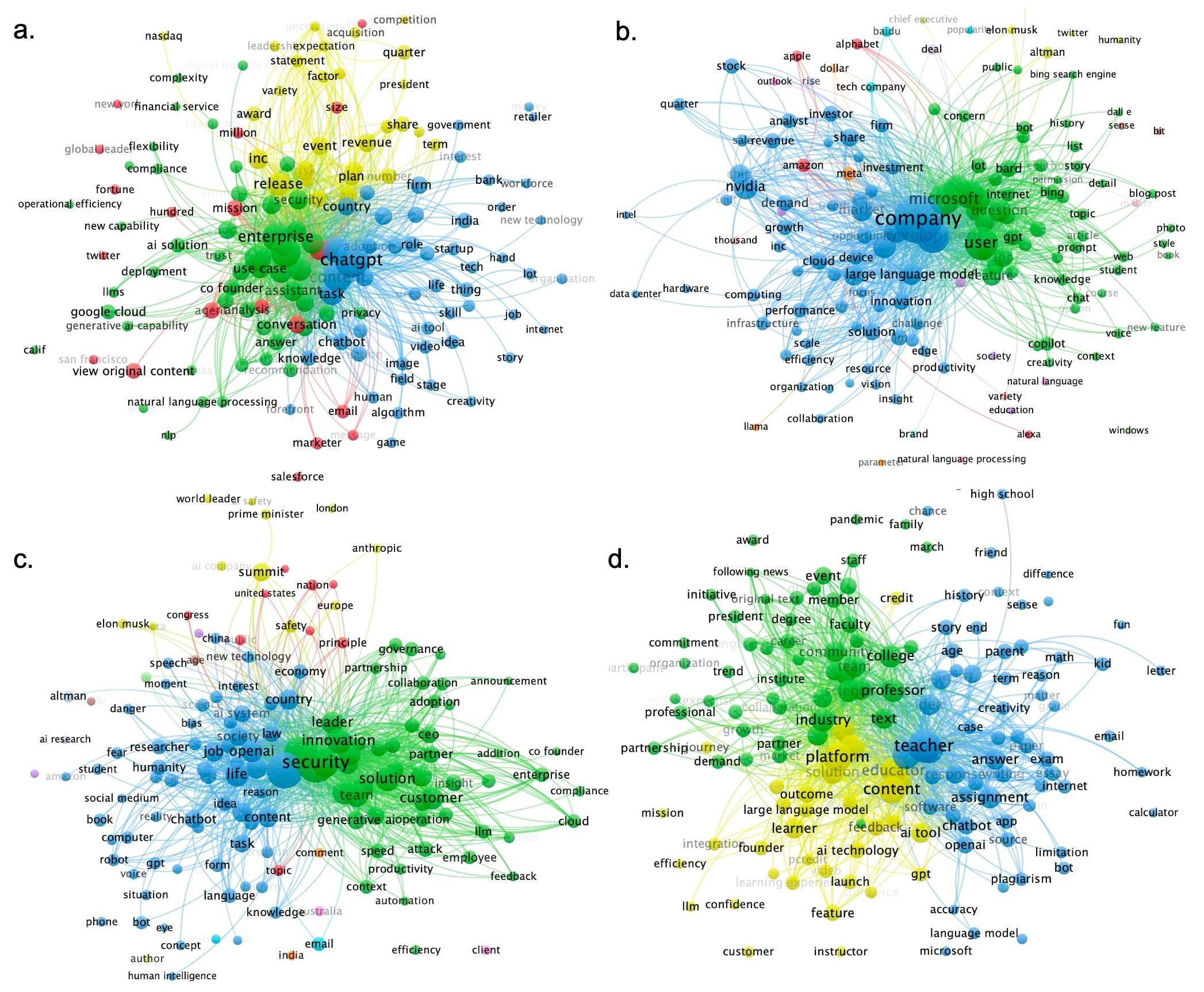}
    \caption{Semantic network for the most discussed four topics covered by news articles, (a) Business, (b) Corporate technological development, (c) Regulation and security, and (d) Education.}
    \label{Fig:semantic}
\end{figure*}

Through semantic network analysis—a method that visualizes the associative relationships between terms in a given discourse \cite{van2014visualizing, jung2020research}—we synthesized the common narratives and terminologies prevalent in the positively-toned news articles within these domains. Utilizing the VOSviewer tool \cite{van2013vosviewer}, we aimed to unearth the frequent sub-themes resonating in optimistic news narratives. The results are presented in Figure~\ref{Fig:semantic}.

Within the semantic network of topic business (Figure \ref{Fig:semantic}(a)), the green cluster centers around generative AI solutions and their applications at an enterprise level. The interplay of terms like ``enterprise'', ``ai solutions'', and ``efficiency'' suggests a focus on the pragmatic application of generative AI in enhancing business processes and reflects an optimistic narrative that appreciates the integration of generative AI, LLMs, and NLP applications in business settings. A separate but equally compelling narrative emerges in the blue cluster: ``startup'', ``technology'', ``workforce'', ``skill'', and ``creativity''. The lexicon of innovation illustrates a dialogue on AI's role in redefining professional landscapes and fostering creative economies. In addition, the yellow-hued cluster emerges on the top of this network, centering on the topics of company revenues, stock shares, and financial statements.

Corporate technological development stands out as another prominent topic in the news discussions, and Figure \ref{Fig:semantic}(b) encapsulates the semantic network for this topic. One significant cluster within this network underscores the contributions of technology giants like ``NVIDIA'', ``Google'', ``Microsoft'', ``Amazon'', ``Meta'', and ``Apple'' to advancing LLM products. This cluster underlines the tech industry's focus on corporate growth, technological innovation, and infrastructure. It also reflects the significance of LLMs and their role in the current tech ecosystem. The other major cluster, represented by the green segment, focuses on how LLM products can elevate knowledge, drive design innovation, and benefit users. Prime examples of such advancements include tools like ``Copilot'', ``Bard'' and ``Bing'' Search. Additionally, prominent figures frequently featured in the news, such as Elon Musk and Sam Altman, are associated with these discussions.

Within the semantic network of regulation and security (Figure \ref{Fig:semantic}(c)), the green cluster emphasizing ``security", ``team", and ``solution", which indicates a discourse centered around cybersecurity and collaborative approaches to safeguarding digital ecosystems. This narrative is complemented by the intertwining of national and global ``governance", hinting at the regulatory and ethical dimensions of AI technology. The blue cluster highlights the discussion of whether to embrace current generative AI trends. These discussions emphasize the ongoing development of generative AI as a prevailing trend. While the progression has revolutionized knowledge and skills, enabling bots to engage in interactive conversations with users, articles within this cluster advocate for regulating AI use for the betterment of humanity. 

Last, the education network reveals three distinct clusters (Figure \ref{Fig:semantic}(d)). The green cluster spotlights the implementation of LLMs within academic institutions and industry, encompassing terms like ``college'', ``institute'', ``faculty'', ``staff'', ``professor'', and ``professional''. This suggests a significant emphasis on AI's role in personalizing learning experiences and managing educational resources.
The blue cluster hints at a narrative surrounding  ``answer'', ``homework'', ``assignment'', ``teacher", ``parent'', and ``creativity'', indicating a focus on the human elements of education, including the impact on students and educators. The overall discourse explores the potential of AI to support or transform traditional educational roles and outcomes. The yellow cluster suggests a more technical discussion is illustrated by the presence of terms like ``large language model", ``gpt", ``outcome", ``platform", and ``tool", suggesting an examination of specific AI tools and their implications for educational technology.


\section{Discussion and Conclusions}

\subsubsection{Key findings}

Our findings provide a comprehensive portrayal of generative AI's reception in global news media based on an in-depth descriptive analysis of news articles after the introduction of BERT in 2018. This paper offers rich insights into the temporal and spatial distribution of topics, sentiment, and substantive themes in major topics of news articles on generative AI.


Temporal patterns reveal variabilities in coverage across key topics—business, corporate technological development, regulation and security, and education—with spikes in articles coinciding with major AI developments and policy discussions, such as the release of Google's Bard and Microsoft's Bing in early 2023, and the White House's AI meeting in May 2023. Spatially, there's a clear delineation of focus, with the US and India leading in business and tech development coverage, whereas the UK, Australia, and Canada are more attuned to regulation and security. Notably, Australia exhibits a unique concentration on education.


Sentiment analysis via a RoBERTa-based model indicates a predominantly neutral to positive media stance, with 66\% of articles being neutral and 28\% positive. News articles on topics like business and corporate technological development show a more positive sentiment, and articles on regulation and security show a more reserved, neutral to negative sentiment, reflective of the critical nature of the discourse in these areas. This finding underscores the intricate nature of global media's portrayal of generative AI's integration into different industries and societies, reflecting the diverse responses it receives in different contexts.


In terms of content, business-related articles focus on the pragmatic application of generative AI at an enterprise level, with a parallel narrative on AI's influence on reshaping professional spheres and fostering creative economies. News about corporate technological development is marked by the contributions of tech giants to LLM products and their implications for user benefit and industry innovation. Regulation and security discussions pivot around cybersecurity, collaborative solutions, and the balancing act between embracing innovation and establishing governance for AI's ethical use. Education-related news coverage unveils three narratives: the integration of AI in academic and professional development, the impact on traditional educational roles, and a technical exploration of AI tools in educational settings. 

\subsubsection{Limitations and opportunities for future work}

Our study presents several limitations, which in turn illuminate potential paths for more comprehensive and detailed research in future endeavors. Firstly, our data collection was restricted to English articles, which could result in biased analysis for those non-English-speaking countries, particularly in the global south. Future efforts could involve collecting news articles in diverse languages, which would provide a more comprehensive global perspective. Limitations in data collection also stem from our access to ProQuest's newsstream databases, which is contingent on our university's subscription policies, as well as from ProQuest's own management and curation of these resources. Although full-text access to the US major dailies newsstream database is not available through our portal, we accessed full-text articles from these outlets via other ProQuest newsstream databases. While we cross-validated these articles by consulting the original news outlets, this approach may not capture the full spectrum of US national news coverage. To address this gap, future research could focus specifically on US national news outlets to ensure a more comprehensive analysis of such news articles.

Secondly, the topic modeling approach presents two limitations. First, the assumption that each document contains only a single topic doesn't always align with the complexity of news articles, potentially leading to suboptimal representations by BERTopic. In addition, we observed instances where clustered topics encompassed news items from different categories. This is possibly because BERTopic clusters rely solely on textual similarity for clustering. Particularly, those clusters with a significant number of news articles are more likely to have diverse topics. To overcome this limitation, future research could explore LLMs, such as integrating GPT-based models into the topic modeling process, allowing for the generation of multiple topics for a single news article.

Thirdly, future work could consider improving sentiment analysis based on more granular units (e.g., paragraph-level or sentence-level) for news articles. Our current sentiment analysis provides a singular classification for each news item, regardless of the potential for conflicting sentiments within an article. This could affect our analysis, especially when a news article covers some topics that might be unrelated to Generative AI. Moreover, the accuracy of sentiment analysis relies on the RoBERTa-base model's capabilities, which weren't specifically trained for sentiment analysis in the news context, possibly leading to inaccuracies. Therefore, another future direction could involve a more nuanced analysis, such as sentiment examination at the paragraph level in news articles.


\section{Acknowledgments}
We would like to express our gratitude to the ProQuest team for providing us with the tutorial on accessing their data.

\bibliography{aaai24}

\end{document}